\documentclass{svproc}
\usepackage{url}

\usepackage[utf8]{inputenc}
\usepackage{graphicx}
\usepackage{tabularx}

\usepackage{float}
\usepackage{booktabs}

\begin{document}
\mainmatter              
\title{Blockchain Network Analysis: A Comparative Study of Decentralized Banks\thanks{ The corresponding author Luyao Zhang is supported by the National Science Foundation China on the project entitled “Trust Mechanism Design on Blockchain: An Interdisciplinary Approach of Game Theory, Reinforcement Learning, and Human-AI Interactions.” (Grant No. 12201266). Yutong Sun is supported by the Summer Research Scholar (SRS) program 2022 under Prof. Luyao Zhang's project entitled "Trust Mechanism Design: Blockchain for Social Good" at Duke Kunshan University. Yufan Zhang and Zichao Chen are supported by the Social Science Divisional Chair’s Discretionary Fund for undergraduate research as the Teaching and Research Assistants of Prof. Luyao Zhang at Duke Kunshan University. Yufan Zhang, Zichao Chen, Yutong Sun, and Luyao Zhang are also with SciEcon CIC, a not-for-profit organization aiming at cultivating interdisciplinary research of both profound insights and practical impacts in the United Kingdom. Yulin Liu is also with Shiku Foundation and Bochsler Finance, Switzerland. We thank the anonymous referees at Computing Conference for their professional and thoughtful comments.}}
\titlerunning{Blockchain Network Analysis: A Comparative Study of Decentralized Banks}  
%
\author{Yufan Zhang\inst{1}
\and Zichao Chen\inst{1} 
\and Yutong Sun\inst{1}
\and Yulin Liu\inst{*}\inst{2}
\and Luyao Zhang\inst{*}\inst{1}}
\authorrunning{Y. Zhang, Z. Chen, Y. Sun, Y. Liu, L. Zhang} %

\institute{Duke Kunshan University, Suzhou, Jiangsu, 215316, China,\\
\email{*corresponding author:lz183@duke.edu \\
Data Science Research Center \& Social Science Division\\
Duke Kunshan University}
\and
SciEcon CIC, London, United Kingdom WC2H 9JQ\\
\email{*corresponding author: yulinzurich@gmail.com}}

\maketitle              

\begin{abstract}
Decentralized finance (DeFi) is known for its unique mechanism design, which applies smart contracts to facilitate peer-to-peer transactions. The decentralized bank is a typical DeFi application. Ideally, a decentralized bank should be decentralized in the transaction. However, many recent studies have found that decentralized banks have not achieved a significant degree of decentralization. This research conducts a comparative study among mainstream decentralized banks. We apply core-periphery network features analysis using the transaction data from four decentralized banks, Liquity, Aave, MakerDao, and Compound. We extract six features and compare the banks' levels of decentralization cross-sectionally. According to the analysis results, we find that: 1) MakerDao and Compound are more decentralized in the transactions than Aave and Liquity. 2) Although decentralized banking transactions are supposed to be decentralized, the data show that four banks have primary external transaction core addresses such as Huobi, Coinbase, and Binance, etc. We also discuss four design features that might affect network decentralization. Our research contributes to the literature at the interface of decentralized finance, financial technology (Fintech), and social network analysis and inspires future protocol designs to live up to the promise of decentralized finance for a truly peer-to-peer transaction network. 

\end{abstract}
\section{Introduction}
Blockchain technology is notable for its security, transparency, and reliability worldwide~\cite{busayatananphon_financial_2022,zhang2022blockchain}. DeFi (Decentralized Finance) is one important blockchain application with over ten billion U.S. dollar market value~\cite{busayatananphon_financial_2022}. According to Werner et al.'s research, DeFi is a peer-to-peer financial system~\cite{werner_sok_2021}. DeFi has great potential to replace traditional centralized finance with the help of blockchain technology by using tamper-proof smart contracts to verify peer-to-peer transactions~\cite{harvey_defi_2021}. Among the various DeFi programs, a class of lending agreements plays a role similar to that of traditional banks. That is, decentralized banks provide lending and borrowing of on-chain assets, facilitated through protocols for loanable funds (PLFs)~\cite{werner_sok_2021}. PLFs can create distributed ledger-based marketplaces for loanable funds of crypto assets by pooling deposited funds in smart contracts~\cite{werner_sok_2021}. Many decentralized banking platforms have emerged in recent years, including Aave, Compound, MakerDao, and Liquity~\cite{busayatananphon_financial_2022}. How do we measure the quality of a decentralized banking platform? According to the study "SoK: Blockchain Decentralization"~\cite{zhang_2022_sok}, decentralized transactions on decentralized banking platforms are important not only because of their financial connotation but also because blockchains with centralized transactions can be easily manipulated by a few individuals~\cite{zhang_2022_sok}, which also threatens blockchain security. Existing literature shows that network characteristics, which are indicators of trading network structures in decentralized markets, significantly affect market outcomes and performance, such as liquidity and volatility~\cite{bovet_evolving_2019},~\cite{motamed_quantitative_2019},~\cite{vallarano_bitcoin_2020}. Furthermore, a more decentralized network can significantly predict higher returns and lower volatility for the associated DeFi tokens~\cite{ao_are_2022}. There is still a lack of sufficient research on blockchain decentralization, and decentralization measurement should include multiple dimensions. 

The paper 'SoK: Blockchain Decentralization' designed a taxonomy to analyze blockchain decentralization in five dimensions: consensus, network, governance, wealth, and transactions, but they found a lack of studies on a transactional perspective~\cite{zhang_2022_sok}. This gap in decentralized banks was filled by a recent study in which Ao et al. (2022)~\cite{ao_are_2022} applied social network analysis to Aave's user blockchain transaction data to capture the degree of decentralization, network dynamics, and economic performance of Aave. They found that the AAVE token transaction network has a distinct core-periphery structure, with multiple network features in a decentralized dynamic state. However, Aave does not represent all decentralized banks. Moreover, the relationship between the mechanism design of DeFi protocols and their degrees of decentralization has not been well studied. Table~\ref{features_table} compares decentralized bank designs in terms of governance, Airdrop, Loan before the deposit, and Stablecoins, where 'Y' is short for 'Yes' and 'N' is short for 'No.' For example, Aave, Compound, and MakerDao have a decentralized autonomous organization (DAO). In contrast, Liquity has an ungoverned protocol that represents a more decentralized mechanism~\cite{liquity}. In addition, Liquity airdrops the native token LQTY to the lenders of the asset pool \cite{liquity}. Before answering how these design mechanisms affect the degree of decentralization of decentralized banks, we need to first compare the degree of decentralization of the platforms and the patterns of their network characteristics. Therefore, our study applies social network analysis~\cite{ao_are_2022} to blockchain transaction network data from leading decentralized banks, including Liquity, Aave, MakerDaok, and Compound and aims to answer the following research questions (RQ).

\begin{itemize}
\item \textbf{RQ1 on network decentralization and dynamics}: How does network decentralization vary over time and across different decentralized bank protocols?
\item \textbf{RQ2 on the core-periphery network}: What are the core components of the transaction network in each decentralized bank? 
\end{itemize}

We complete the core-periphery analysis and characterization of the transaction network using the LIP algorithm~\cite{LIP_2011_a}. Our research successfully compares the network features and transaction decentralization of four major decentralized bank protocols. Due to the introduction of multiple platforms in our comparison, the quantity of data in our study is tens of times larger than that in the previous study~\cite{ao_are_2022}. To solve the technical issue, we introduce the LIP algorithm, a faster core-peripheral analysis algorithm, in our study~\cite{LIP_2011_a}.

Our results (R) reveal that
\begin{itemize}
\item \textbf{R1}: MakerDao and Compound are more dispersed in transactions than AAVE and Liquity. 
\item \textbf{R2}: The largest externally owned address cores for LQTY, LUSD, AAVE, Dai, and COMP are centralized exchanges such as Huobi, Coin base, and Binance.
\end{itemize}
The rest of the paper is organized as follows. Section 1.1 addresses the related literature. Section 2 introduces the data and methods. Section 3 presents the results. Section 4 concludes and discusses future research. 
\begin{figure}[h]
  \centering
  \includegraphics[width=0.85\textwidth]{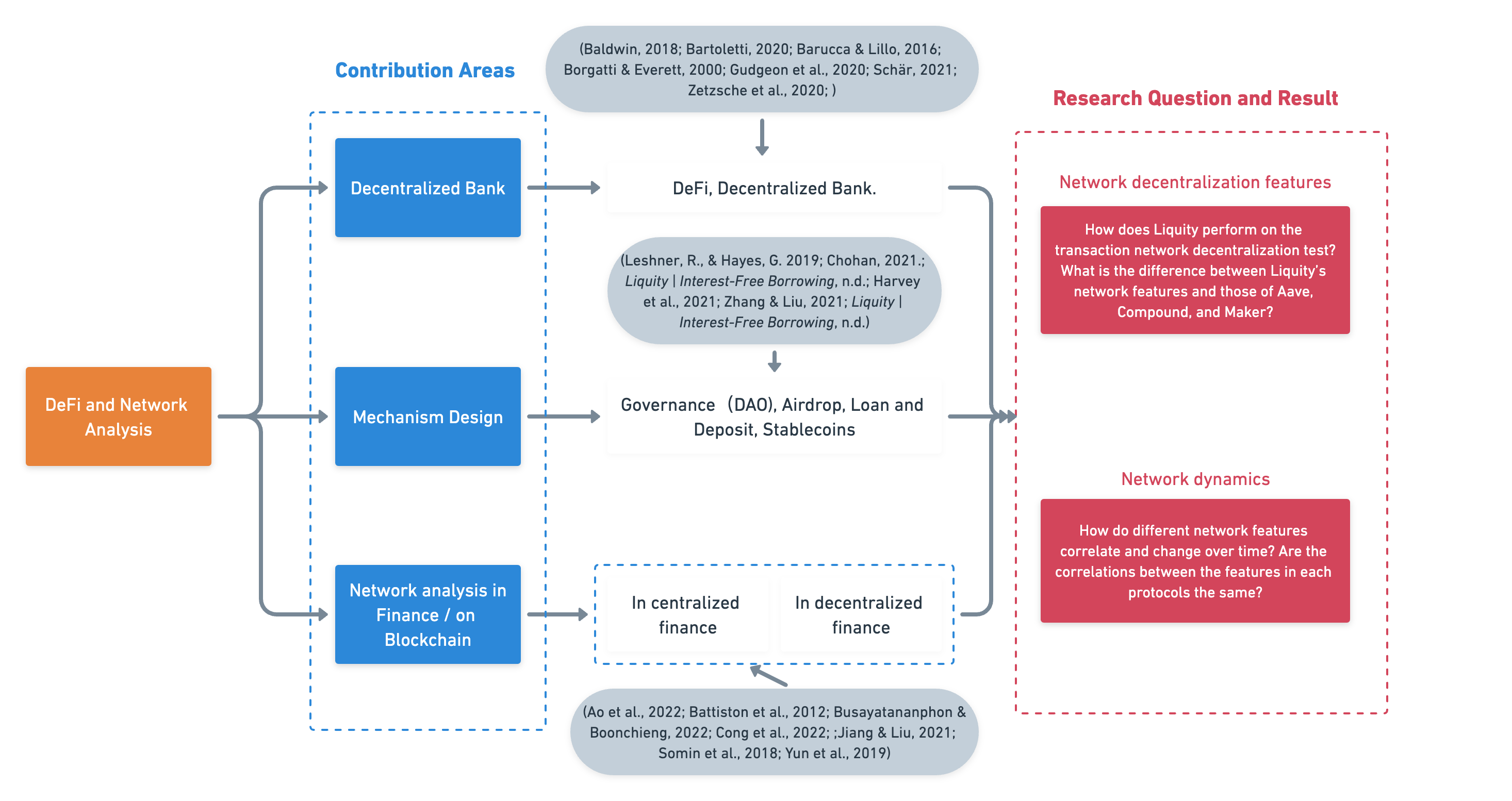}
  \caption{Literature review flowchart}
  \label{Literature}
\end{figure}

\begin{table*}
\centering
\caption{Mechanism design comparison between four decentralized banks}
\begin{tabular}{l||c|c|c|c}
\hline
Platforms & Governance & Airdrop & Loan before the deposit & Stablecoins \\
\hline
Compound & Y & Y & N & N\\
Aave & Y & N & N & N\\
MakerDao & Y & N & Y & Y, Dai\\
Liquity & N & Y & Y & Y, LUSD\\
\hline
\end{tabular}
\label{features_table}
\end{table*}

\subsection{Literature}
As in Figure~\ref{Literature}, our research contributes to four lines of literature including decentralized finance (DeFi), the mechanism design of decentralized bank network analysis in finance, and blockchain network analysis.

\subsubsection{Decentralized Finance}
Our research contributes to the decentralized bank literature in the DeFi area. DeFi, decentralized finance, is one of the most discussed emerging technological evolutions in global finance~\cite{zetzsche_decentralized_2020}. DeFi refers to an alternative financial infrastructure built on top of the Ethereum blockchain~\cite{schar_decentralized_2021}. It uses smart contracts on blockchain to replicate the existing financial services in a more open, interoperable, and transparent way~\cite{schar_decentralized_2021}. DeFi brings a brand new trust mechanism in this digital era, which implies a move from trust in banks or states to trust in algorithms and encryption software, according to research~\cite{baldwin_digital_2018}. Among decentralized finance, the decentralized banks take up a large area~\cite{zhang_2022_sok}. According to Ao et al.(2022)~\cite{ao_are_2022}, decentralized banks differ from centralized banks in two aspects: 1) they replace centralized credit assessments with coded collateral evaluation~\cite{gudgeon_defi_2020}, and 2) they employ smart contracts to execute asset management automatically~\cite{bartoletti_smart_2020}. We introduce the network analysis to several decentralized bank transaction networks~\cite{ao_are_2022}~\cite{barucca_disentangling_2016}~\cite{borgatti_models_2000}, including Aave, Liquity, Compound, and MakerDao. Our research contributes to decentralized banks by innovatively comparing the decentralized network features across several decentralized banks.

\subsubsection{DAO, Airdrop, Stablecoins, Loan and Deposit Mechanism Design}

Centralized exchanges, such as Bitfines or Poloniex, are trust-based systems and have many limitations. In contrast, decentralized banks facilitate a much more convenient loan experience, according to the Compound white paper, a famous decentralized bank established in 2016~\cite{Compound}. Decentralized banks have brought customers a brand new mechanism for lending, borrowing, and depositing. In terms of platform design, different decentralized banks have different mechanisms. These mechanisms are divided into four aspects: Governance, Airdrop, Loan and Deposit, and Stablecoins. In Table~\ref{features_table}, we provide the different mechanisms used by the different platforms. In the governance mechanism, decentralized banks such as Compound, Aave, and MakerDao all have a decentralized autonomous organization (DAO). Token holders can use their tokens to participate in community governance~\cite{chohan_decentralized_2017}. However, Liquity designs a completely autonomous DeFi protocol without a DAO or centralized governance. This innovative mechanism may help it to be more decentralized and transparent~\cite{liquity}. Second, in the airdrop process, the incentive mechanism, Compound, and Liquity deliver rewards tokens to the users for liquidity mining~\cite{liquity}~\cite{Compound}. The incentive airdrop reward may help the decentralized bank obtain better liquidity and attract more users~\cite{yin_liquidity_2021}. In a traditional depositing scheme, people can lend an asset only if someone deposits it in a liquid pool~\cite{jakab_banks_2015}. However, Liquity and MakerDao devised a new mechanism that allows customers to borrow an asset before any deposits~\cite{liquity}. This loan before the deposit mechanism  of MakerDao and Liquity also introduces another stablecoin design. Stablecoins are one type of decentralized finance application~\cite{harvey_defi_2021} intended to remedy cryptocurrencies' excess volatility. During stablecoin development, there have been 3 main periods: fiat-backed, crypto-backed, and algorithmic stablecoins~\cite{baur_crypto_2021}. A major role of stablecoins is to provide security and stability to investors. Compared with volatile cryptocurrencies such as Bitcoin, research has found that stablecoins act as a safe haven for bitcoin~\cite{Compound}. For stablecoins, MakerDao introduced the Dai stablecoin and Liquity introduced the stablecoin LUSD stablecoin~\cite{liquity}. Liquity designed a hard and soft peg mechanism for the LUSD stablecoin. Our paper contributes to the decentralized bank mechanism design, including Governance, Airdrop, Loan and Deposit, and Stablecoins, by exploring and analyzing the potential influence of these innovative mechanism designs on the transaction networks of four decentralized banks.

\subsubsection{Network Analysis in Finance}
Applying social network analysis to financial markets became popular after the financial crisis from 2008 to 2009~\cite{ao_are_2022}. Many studies have found that it is important to explore and evaluate financial network structures to identify systemic risks. Banks that are too centralized, for example, may cause a chain reaction in which their failure may destroy the wider financial system \cite{battiston_debtrank_2012} \cite{yun_too_2019}. Cong et al.'s (2022)~\cite{cong_inclusion_2022} research introduced the network analysis method to the decentralized financial system and successfully analyzed the Ethereum financial network. By introducing network analysis methods for several decentralized banks and comparing the results, our paper further demonstrates the transaction networks and possible influences of decentralized banks, contributing to the field of the decentralized finance field.

\subsubsection{Network Analysis on Blockchain}
As a newly emerged and highly concerned field, blockchain is undergoing very rapid development~\cite{busayatananphon_financial_2022}. A blockchain is not only a distributed ledger but a network of transactions~\cite{somin_network_2018}. Each account address in the blockchain can be thought of as a network node. Somin et al. (2018)~\cite{somin_network_2018} started their research on network analysis research of ERC20 tokens rending on the Ethereum blockchain and demonstrated the strong power-law properties, richer get richer, in the network. Then Jiang and Liu (2021)~\cite{jiang_cryptokitties_2021} spread the network analysis to the NFT area, where they analyze the CryptoKitties' transaction network. Cong et al (2022)~\cite{cong_inclusion_2022} provide a network analysis of decentralized finance network analysis on Ethereum. Ao et al. (2022) analyze the Aave decentralized bank transaction network in more detail using the core-periphery method and network feature analysis~\cite{ao_are_2022}. Our paper contributes to the field of blockchain network analysis field, delving further into existing blockchain-based analysis and exploring the transaction network of decentralized banks in a more detailed manner.

\section{Data and Methods}
\label{Data}
\textbf{Data and code availability} We made our code and data open-source, available on GitHub\footnote{https://github.com/SciEcon/Blockchain-Network-Analysis}.

\begin{figure}[h]
  \centering
  \includegraphics[width=0.9\textwidth]{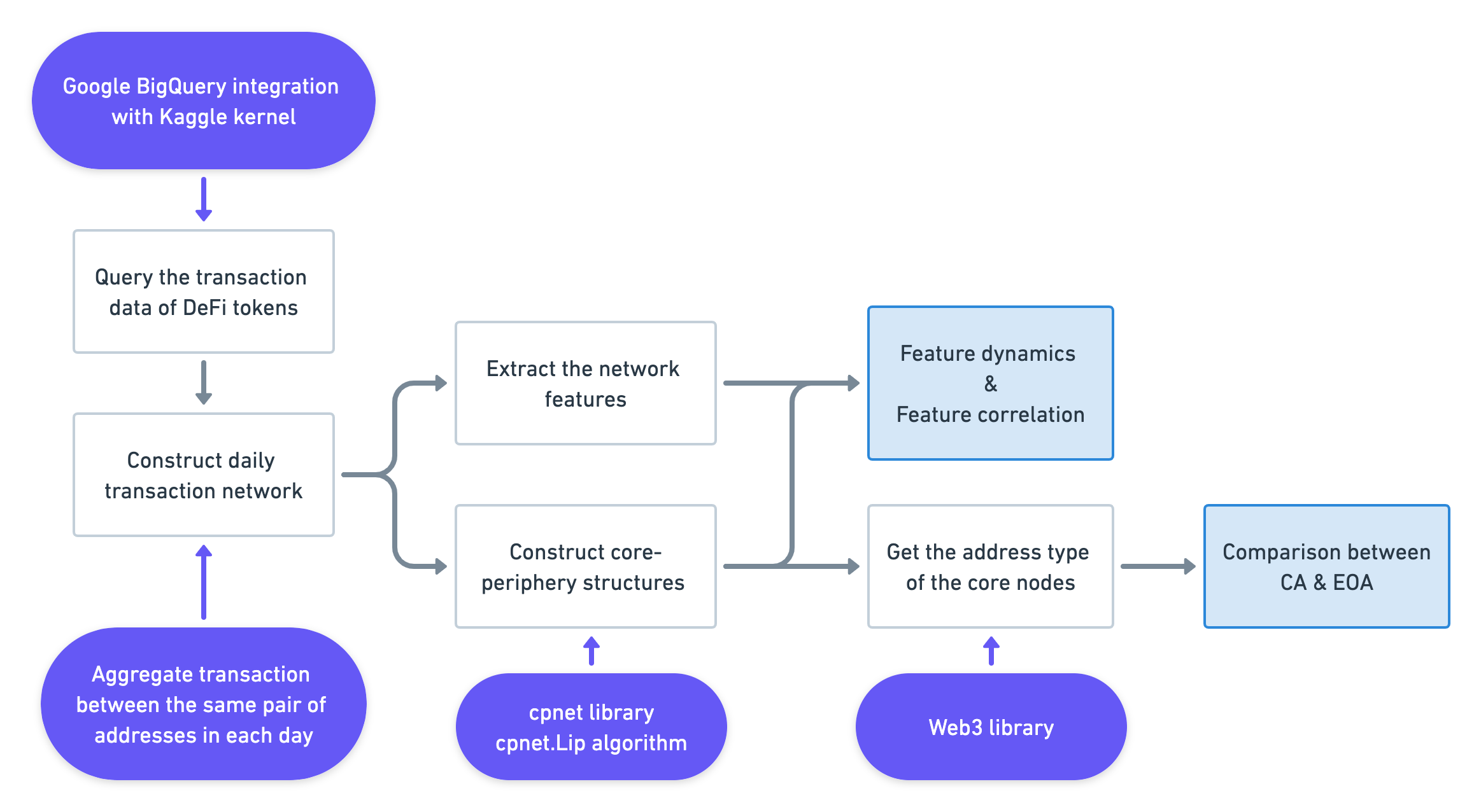}
  \caption{Blockchain network analysis methodology flowchart}
  \label{MethodFlow}
\end{figure}

Figure~\ref{MethodFlow} depicts the data science pipeline of our study. Our study has higher automation and computational efficiency than earlier studies~\cite{ao_are_2022}, enabling cross-sectional comparisons. 

\subsection{Data Source and Preprocessing}
The data for blockchain network analysis are derived from the transaction records of 5 tokens from 4 decentralized finance protocols: LUSD and LQTY of Liquity~\cite{liquity}, AAVE of Aave~\cite{aave}, Dai of MakerDao~\cite{makerdao}, and COMP of Compound~\cite{Compound}.
These transaction records were obtained from BigQuery dataset of the Ethereum blockchain via the BigQuery integration with the Kaggle kernel~\cite{ethereum}. The launch dates for the DeFi protocols are summarized in Table~\ref{data_table}. Specifically, the data cover the transaction records from the genesis dates of each token to July 12, 2022. The transactions whose from\_address or to\_address is the Ethereum null address, which is often associated with token-related events such as genesis, mint, or burns \cite{etherscanio_2019_ethereum}, were filtered out. We also summarize the total transaction value involved and the number of addresses in Table~\ref{data_table}. Figures~\ref{LUSD_1},~\ref{LQTY_1},~\ref{AAVE_1},~\ref{Dai_1} and~\ref{COMP_1} in the Appendix visualize the change in daily transaction value and the number of addresses over time. In addition, we developed undirected daily transaction networks in which the nodes represent Ethereum addresses, and the edges represent the entire daily transaction volume between two addresses, weighted by the transaction values. In other words, we aggregate the transaction values between two addresses without regard to the direction so that the transaction between two addresses will not be calculated repeatedly in the core-periphery structure analysis. 

\label{Data source and preprocessing}

\begin{table*}[!htbp]
\centering
\caption{Queried data of the DeFi tokens for blockchain network analysis.}
\resizebox{\textwidth}{!}{
\begin{tabular}{llllll}
\hline
Token & DeFi Protocol & Genesis Date & Duration (Day) & Total Transaction Value (Wei) & Number of Addresses \\ \hline
LUSD & Liquity  & 2021-04-05 & 464 & $2.823 \times 10^{28}$  & 12,249    \\
LQTY & Liquity  & 2021-04-05 & 464 & $4.147 \times 10^{26}$  & 19,009    \\
AAVE & Aave     & 2020-10-02 & 649 & $3.511 \times 10^{26}$   & 371,122   \\
Dai  & MakerDao & 2019-11-13 & 968 & $1.295 \times 10^{30}$  & 1,769,138 \\
COMP & Compound & 2020-03-04 & 793 & $1.4823 \times 10^{26}$ & 567,133   \\ \hline
\end{tabular}}
\label{data_table}
\end{table*}

\subsection{Network Feature Extraction}
We extracted 4 network features for all the daily transaction networks built as described in Section~\ref{Data source and preprocessing}. In detail, the network features include \textbf{the number of components}, \textbf{the relative size of the largest component}, \textbf{the modularity score} \cite{newman_2006_modularity}, \textbf{and the standard deviation in the degree centrality}. These network features are computed using the Python NetworkX \cite{networkx} algorithms. According to the conceptual framework, the network features can characterize the difference between more centralized and more decentralized networks and therefore quantify the decentralization of a specific transaction network. For instance, in a 'more centralized' network, in which more vertices are connected to several central vertices, the relative size of the largest component will become larger compared to the decentralized ones.

\subsection{Core-periphery Structure Detection}
There are two more features for blockchain network analysis that require detecting the core-periphery structure of the daily transaction network. The core-periphery structure refers to a fundamental network pattern that categorizes network nodes \cite{doi:10.1126/sciadv.abc9800}. Specifically, the network nodes are classified into two categories: “core” nodes, which are densely connected, and “periphery” nodes, which are weakly connected. Transaction networks with a more significant core-periphery structure are more decentralized than those with a less significant structure \cite{ao_are_2022}. There are several algorithms for detecting the core-periphery structure of a given network \cite{BORGATTI2000375} \cite{LIP_2011_a}, \cite{PhysRevE.96.052313} \cite{BOYD2010125} \cite{cucuringu2016detection}. By evaluating the statistical significance of the core-periphery structure in a transaction network, we can construct two additional features, \textbf{the number of detected core nodes} and \textbf{the average degree of the core nodes}, for the blockchain network analysis. The core-periphery structure construction and statistical analysis are conducted using the LIP algorithm \cite{LIP_2011_a} in the Python cpnet library \cite{kojaku_python_2022}. Together with the four extracted features, the two new network features are formally defined in Table~\ref{definition}.

\begin{table}[!htbp]
\caption{Definition of the extracted network. ↓ means the lower the less decentralized, and ↑ means the more decentralized.}
\begin{tabular}{p{0.3\linewidth} p{0.7\linewidth}}
\hline
Feature                                 & Definition                                                                                                                    \\ \hline
The number of components ↑                        & The number of connected subnetworks that are not part of any larger connected network given a transaction network.           \\
The relative size of the largest component ↓      & The number of nodes in the largest component divided by the total number of nodes.                                          \\
Modularity score ↑                            & The fraction of the edges that fall within the given groups minus the expected fraction if edges were randomly distributed. \\
The standard deviation of the degree centrality ↓ & The standard deviation of the degree of each node in a given transaction network.                                           \\
The number of detected core nodes ↑               & The number of nodes that are detected as “core” by the LIP algorithm.                                                       \\
The average degree of the core nodes ↓            & The average degree of the nodes that are detected as “core” by the LIP algorithm.                                           \\ \hline
\end{tabular}
\label{definition}
\end{table}

\section{Results}

\subsection{Network Feature Dynamics and Correlation}

\begin{figure*}[!htbp]
  \centering
  \includegraphics[width=\textwidth]{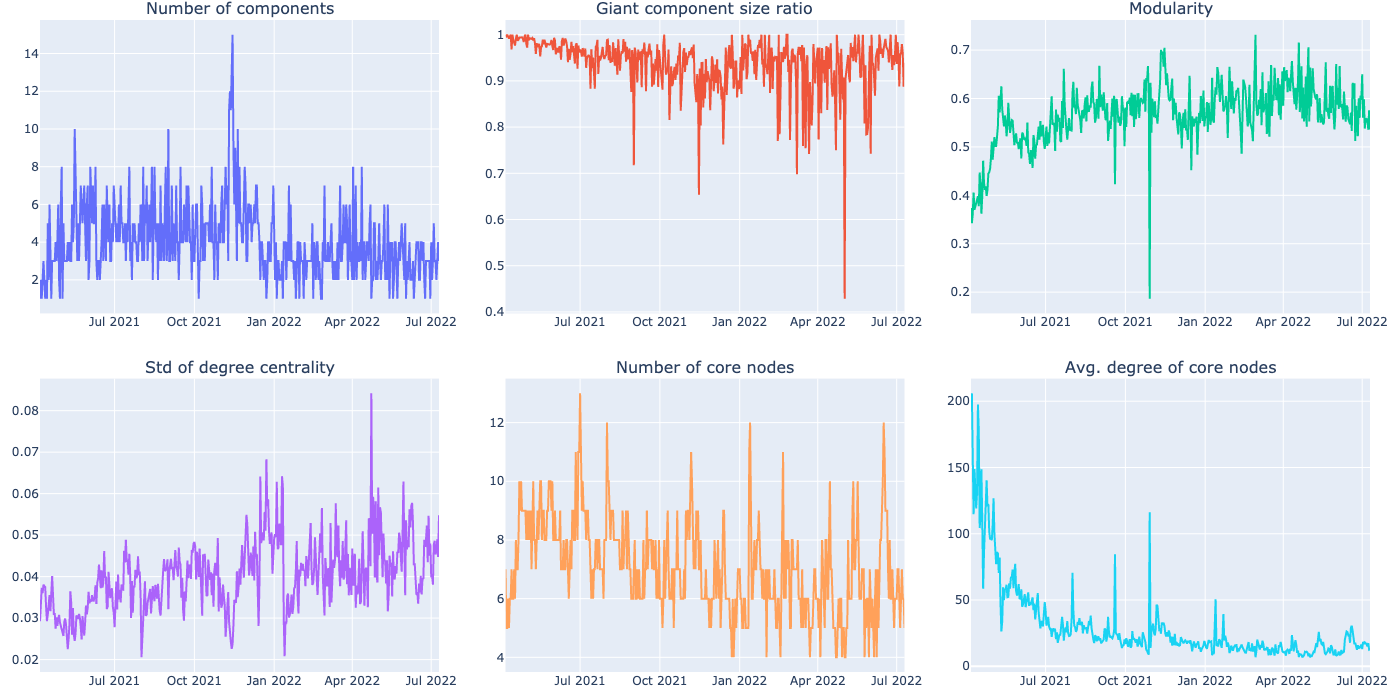} 
  \caption{Time-series plots of network features of the AAVE token.}
  \label{AAVE_dynamics}
  \includegraphics[width=\textwidth]{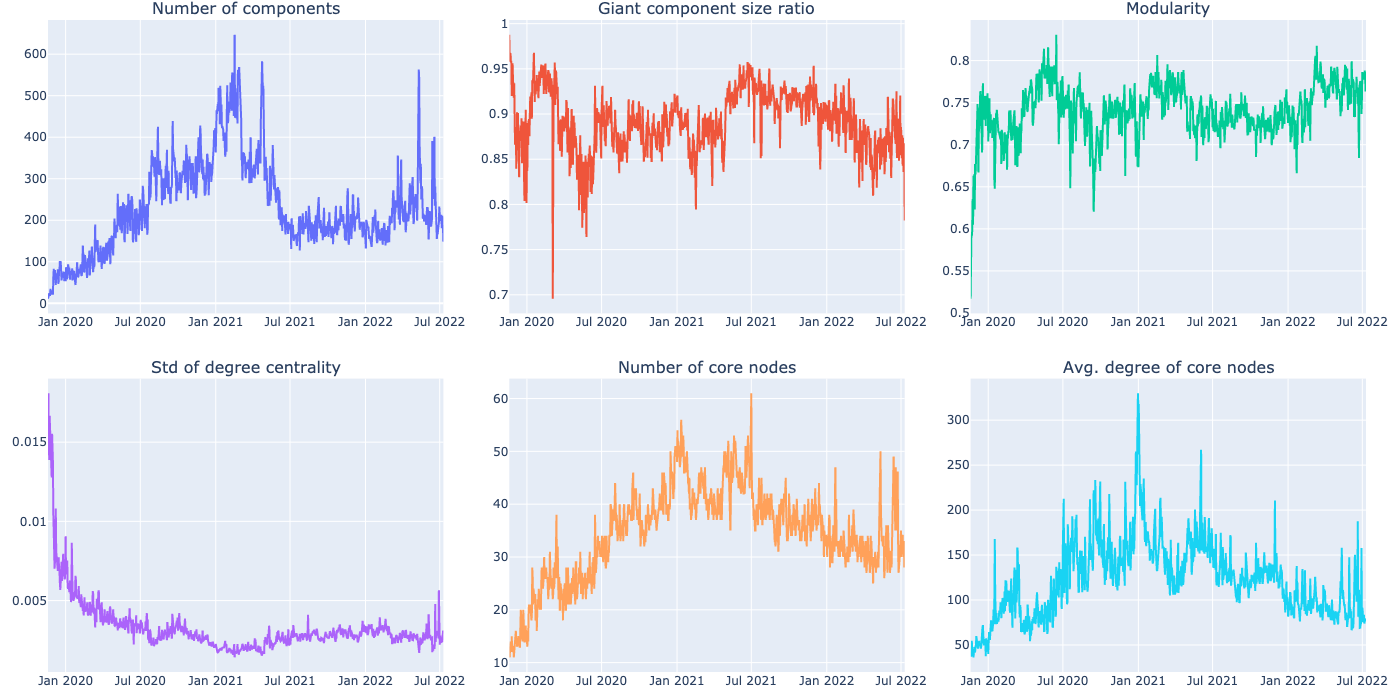} 
  \caption{Time-series plots of network features of the Dai token.}
  \label{Dai_dynamics}
  \includegraphics[width=\textwidth]{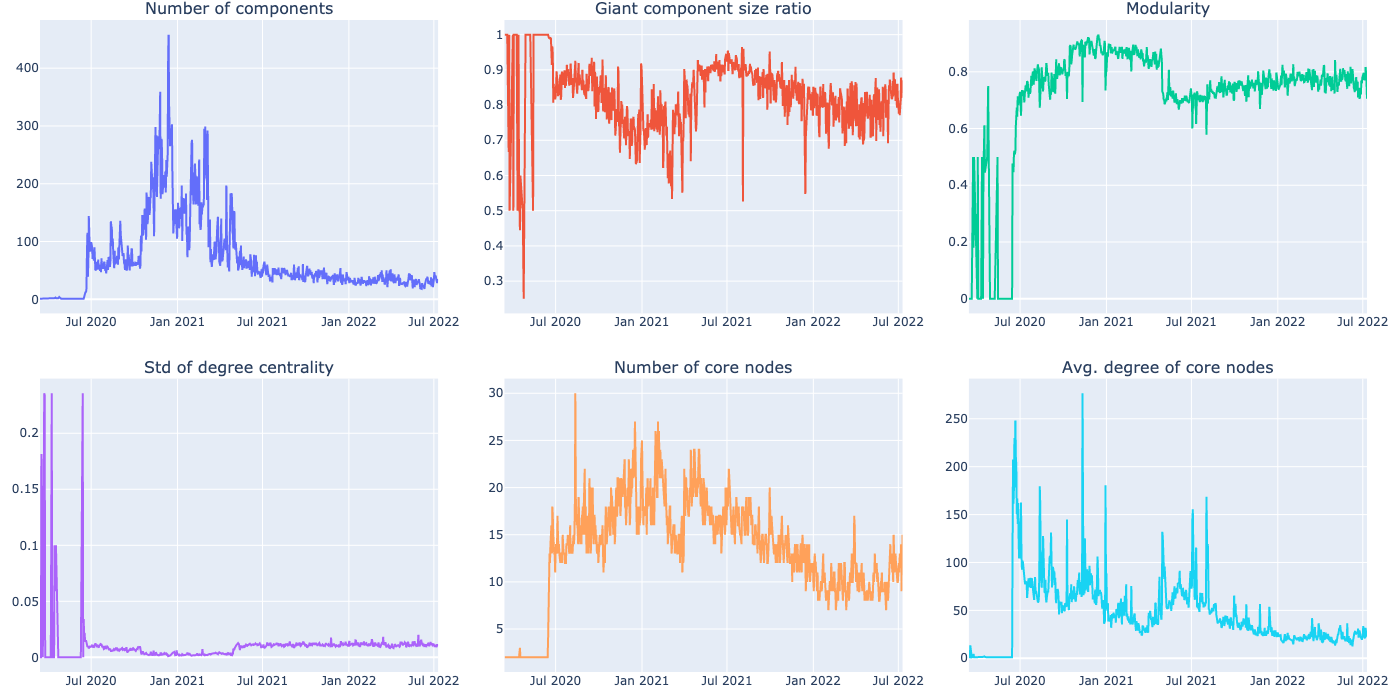} 
  \caption{Time-series plots of network features of the COMP token.}
  \label{COMP_dynamics}
\end{figure*}

\begin{figure*}[!htbp]
  \centering
  \includegraphics[width=\textwidth]{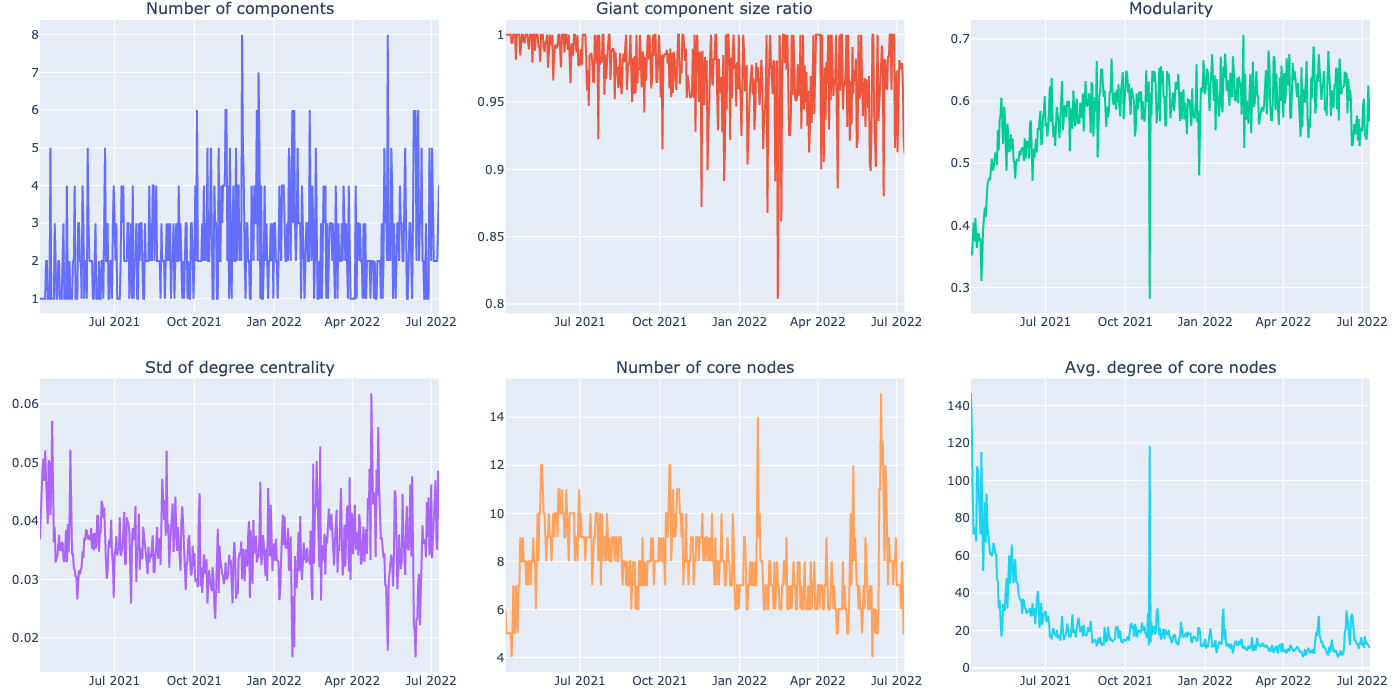} 
  \caption{Time-series plots of network features of the LUSD token.}
  \label{dynamics_LUSD}
  \includegraphics[width=\textwidth]{img/LQTY/fig3.png} 
  \caption{Time-series plots of network features of the LQTY token.}
  \label{dynamics_LQTY}
\end{figure*}

Given the calculated network features for each transaction network of each token, we explore how the decentralization of the transaction network for the five DeFi currencies evolves over time using the six network features introduced in Section~\ref{Data}. The time series plots of the network features for each token are illustrated in Figures~\ref{AAVE_dynamics},~\ref{Dai_dynamics},~ \ref{COMP_dynamics},~\ref{dynamics_LUSD} and~\ref{dynamics_LQTY}. 
By examining the relationship between these dynamic features and the degree of decentralization, we first validated the conceptual framework introduced in Table~\ref{definition}. Horizontally comparing the dynamic features of these four platforms reveals that COMP from Compound and Dai from MakerDao have a substantially higher degree of decentralization than AAVE from Aave and LQTY/LUSD from Liquity.

In addition to calculating the network dynamics, we measured the correlations between network characteristics to better highlight their relationship with network decentralization. Figures~\ref{cor_LUSD}, \ref{cor_LQTY}, \ref{cor_AAVE}, \ref{cor_Dai} and \ref{cor_COMP} in the Appendix depict the feature correlation for LUSD, LQTY, AAVE, Dai, and COMP tokens respectively. Through the correlation heatmap, we can determine the degree of correlation between each platform's network properties. The stronger the correlation, the darker the square. The greater the degree of correlation between network variables moving in the same direction, the more effective and rational the network analysis for that bank's network. AAVE and Dai have a larger correlation degree in the heatmap comparison, but LQTY and LUSD of the platform have a comparatively low correlation degree. This may suggest that the existing trading network is immature and that its features are obscure.

In the horizontal comparison of these five tokens' network features, we selected the following aspects of  the significant comparison results.

\subsubsection{Number of components}
In this comparison, we discovered that both of the earliest established platforms, MakerDao and Compound, experienced a peak in the number of components around January 2021, followed by a continuous decline. Next, the graph demonstrates that there is a significant difference in the number of components between AAVE and LQTY and that the quantity of AAVE is much greater than that of LQTY. This indicates that Liquity's trading network may be more straightforward than competing platforms.

\subsubsection{Modularity score}
The smaller the modularity score, the more centralized the market. In the horizontal comparison, both LQTY and LUSD had lower values than the two earlier platforms, Compound and MakerDao. This may suggest that Liquity's trading network is less decentralized than the two previous platforms.

\subsubsection{Standard of degree centrality}
We discover that LQTY and LUSD have a greater standard degree of centrality than other platforms based on these data. Both Compound and MakerDao, two older decentralized banks, have rather poor scores in this category. The values of LQTY and LUSD likewise exhibit an upward trend. The lower the value, the less centralization there is. This conclusion is identical to the modularity finding, and it matches the results of other network dynamic properties as well. This circumstance suggests that the Liquity platform is currently more centralized than other platforms.

\subsection{Core-periphery Structure Comparison Between Contract Addresses and Externally Owned Addresses}

\begin{figure*}[!htbp]
  \centering
  \includegraphics[width=\textwidth]{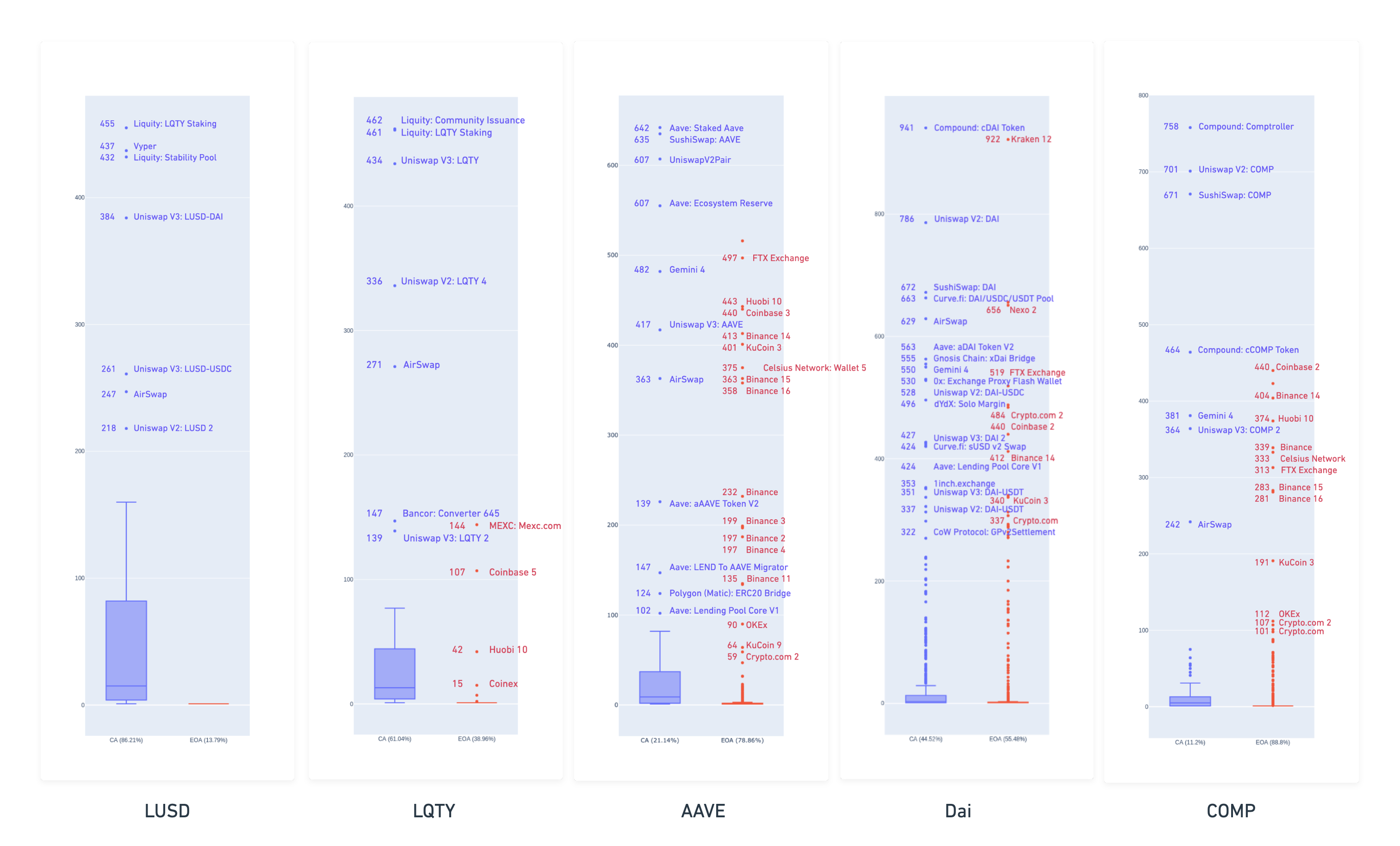} 
  \caption{Distribution of the number of core days for EOAs and CAs of the five tokens with the annotated address information of the outliers.}
  \label{address_type}
\end{figure*}

In addition to the network feature analysis, we conducted further exploration of the detected core-periphery structure for the transaction networks of each token. As introduced in Section~\ref{Data}, we conducted the core-periphery structure analysis on the daily transaction networks for the DeFi tokens, LUSD, LQTY, AAVE, Dai, and COMP. 
Using the Python Web3 package \cite{merriam_introduction}, we subsequently queried the real types of addresses detected as core nodes.
To further investigate the decentralization of a DeFi token, we further contrasted the core-periphery structure study results in terms of the address types.
On the Ethereum blockchain, there are two types of addresses: contract addresses (CA) and externally owned addresses (EOA). The former represents the executable smart contract on the blockchain, whereas the latter consists of user accounts. We extracted the unique addresses that were detected as core nodes for at least one day within the time span of the data source. Figure \ref{address_type} demonstrates the distribution of the days for them being core nodes in terms of the address type. Moreover, we investigated the detailed address information via Etherscan.io \cite{etherscanio_2019_ethereum} of the outlier addresses of the distribution, which are also annotated in Figure \ref{address_type}.

We observe that the CA outliers with the most core days are the token contracts created by the DeFi protocol developer. For instance, the CA with the highest number of core days in the AAVE transaction network is the \textit{Aave: Staked Aave} (642 days), which is the token contract of the AAVE token. Other outliers of CA are the decentralized cryptocurrency exchanges built on Ethereum using smart contracts. For instance,\textit{Uniswap} \cite{uniswap} is one of the automated liquidity protocols powered by smart constants that exist as outliers in all five token transaction networks, which enables peer-to-peer market making. Another decentralized cryptocurrency exchange that exists as the CA outlier for all five tokens is \textit{AirSwap} \cite{airswap}, which can also archive peer-to-peer trading of Ethereum tokens. The outliers among EOAs are mostly centralized cryptocurrency exchanges. The most obvious examples are \textit{Coinbase} \cite{coinbase} and \textit{Binance} \cite{binance}, both of which are famous exchanges where users can trade cryptocurrencies. Given the vast variety of trading tools and supporting services for users to earn interest \cite{binance}, centralized cryptocurrency exchanges with a high volume of transactions have gained immense appeal, where a large number of transactions occur on these EOAs. However, the centralized exchanges bring high centralization” to the decentralized bank transaction networks. Table~\ref{list_date} summarizes the first list date for the four DeFi protocols.

\begin{table}[!htbp]
\centering
\caption{The first date that the DeFi protocols were first listed on the exchanges, Coinbase and Binance}
\begin{tabular}{ccc}
\hline
         & Coinbase          & Binance          \\ \hline
Liquity  & January 12, 2022  & Not yet          \\
Aave     & December 14, 2020 & October 16, 2020 \\
Compound & June 23, 2020     & June 25, 2020    \\
DAO & May 23, 2019      & July 23, 2020    \\ \hline
\end{tabular}
\label{list_date}
\end{table}

\section{Conclusion and Future Research}
According to Cong et al. (2022)~\cite{cong_inclusion_2022}, the degree of decentralization and the stability of the trading network are both important factors in building trust for decentralized banks and increasing the inclusiveness of the decentralized bank platform. We, therefore, conducted a comparative study of four major decentralized banks including Liquity, Aave, MakerDao, and Compound, evaluating transaction decentralization using social network analysis. We made two major findings: first, the largest externally owned address cores for LQTY, LUSD, AAVE, Dai, and COMP mainly include exchanges such as Huobi, Coin base, and Binance, second, MakerDao and Compound are more decentralized in trading than AAVE and Liquity. 

Future research can further study the connection between protocol designs and the decentralization level. For example, the higher level of centralization on LQTY and LUSD may be due to three reasons. First, as the Liquity platform has not been established for a long time, there may be fewer users on the platform. Second, as LQTY has not yet been listed on some exchanges, such as Binance, the token may be less well known. This may lead to distrust of the Liquity platform by other decentralized bank participants, resulting in fewer addresses participating in the trading network. Third, Liquity is designed as a non-governance system, which may leave the platform without royalty users actively interacting in the network. How would the internal design features of governance, airdrop, loan before deposits, and stablecoins and the external events such as blockchain mechanisms upgrade~\cite{zhang2023understand,liu2022} affect network decentralization and other desired properties~\cite{zhang2023design}? Our study provides a direction for future exploration of transaction network analysis and mechanism design for decentralized banks.

\bibliographystyle{spmpsci}
\bibliography{citation}

\appendix
\section{Appendix}

\begin{figure}[!htbp]
  \centering
  \includegraphics[width=0.7\linewidth]{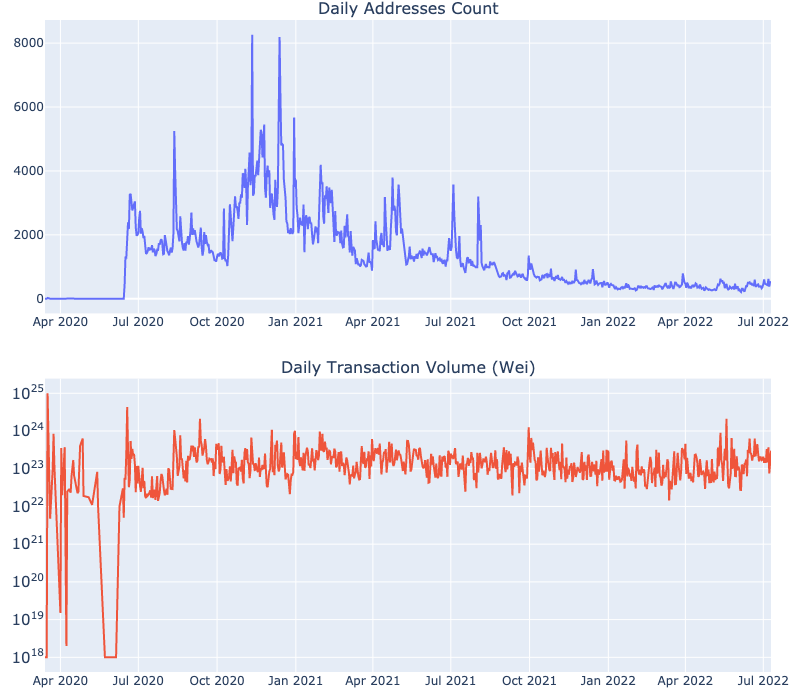} 
  \caption{Time series plots of the daily transaction value (Wei) and the number of addresses of the COMP token.}
  \label{COMP_1}
\end{figure}

\begin{figure}[!htbp]
  \centering
  \includegraphics[width=0.7\linewidth]{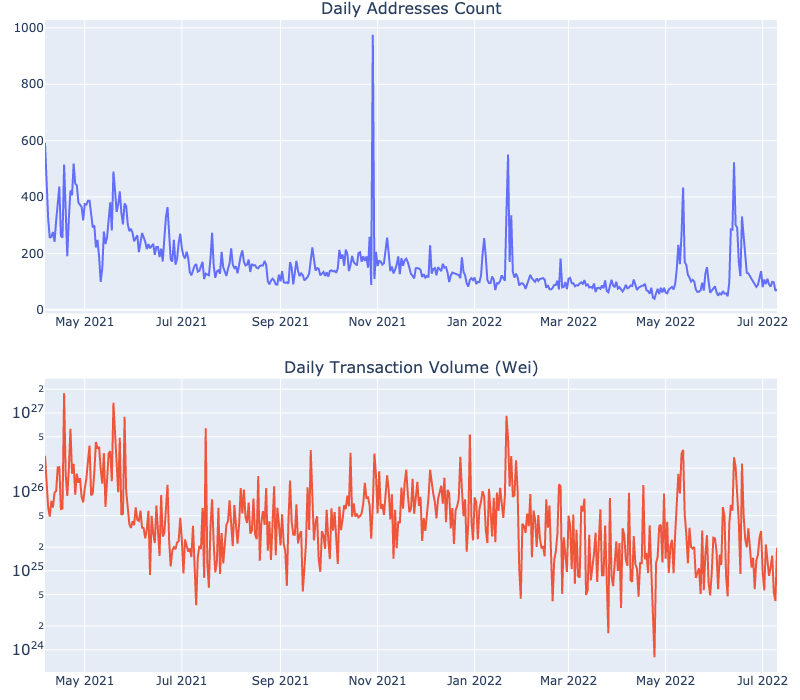} 
  \caption{Time series plots of the daily transaction value (Wei) and the number of addresses of the LUSD token.}
  \label{LUSD_1}
  \includegraphics[width=0.7\linewidth]{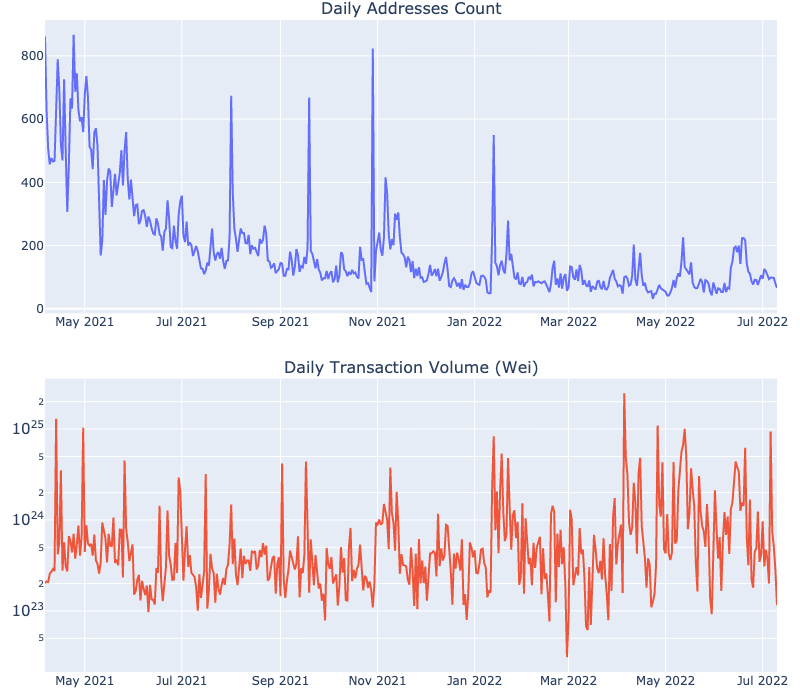} 
  \caption{Time series plots of the daily transaction value (Wei) and the number of addresses of the LQTY token.}
  \label{LQTY_1}
\end{figure}

\begin{figure}[!htbp]
  \centering
  \includegraphics[width=0.7\linewidth]{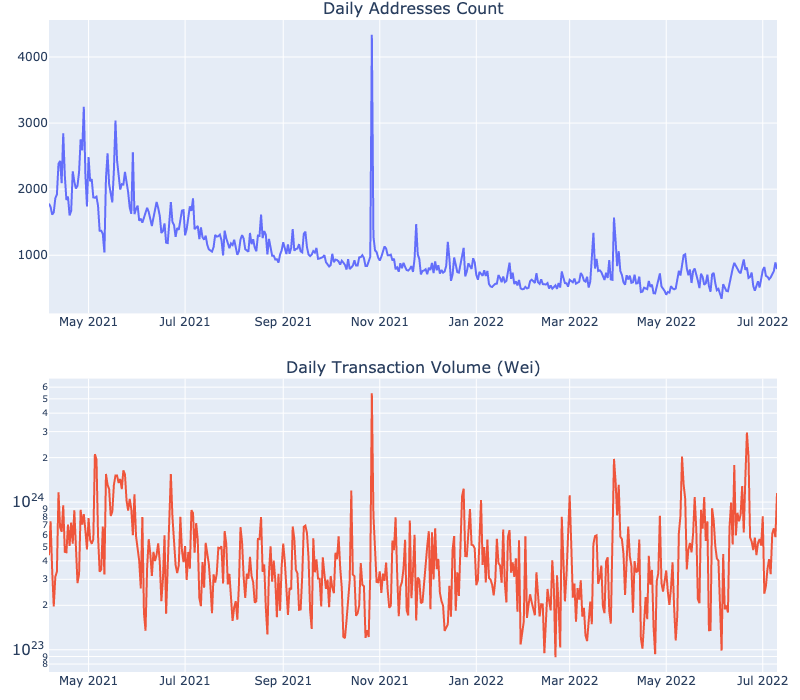} 
  \caption{Time series plots of the daily transaction value (Wei) and the number of addresses of the AAVE token.}
  \label{AAVE_1}
\end{figure}

\begin{figure}[!htbp]
  \centering
  \includegraphics[width=0.7\linewidth]{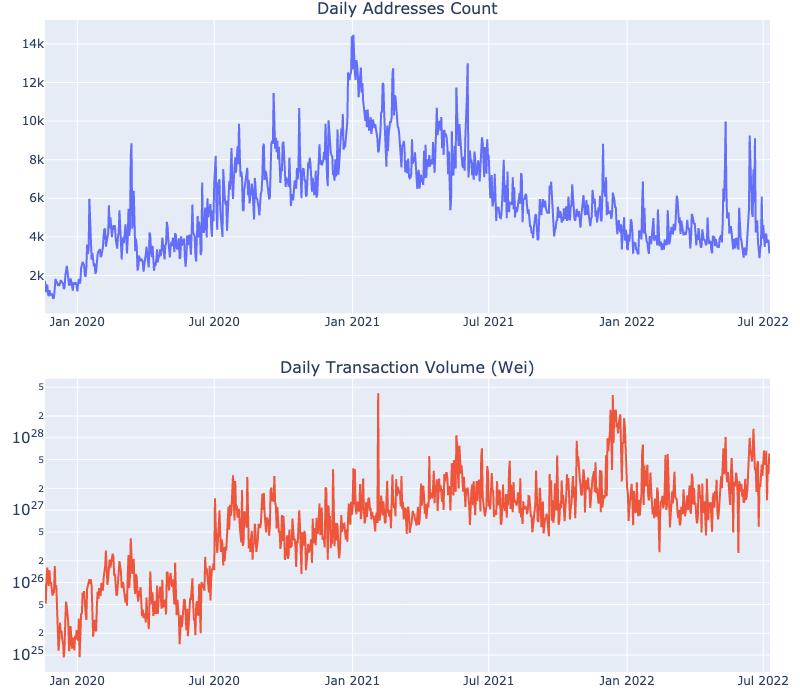} 
  \caption{Time series plots of the daily transaction value (Wei) and the number of addresses of the Dai token.}
  \label{Dai_1}
\end{figure}

\begin{figure*}[!htbp]
  \centering
  \includegraphics[width=\textwidth]{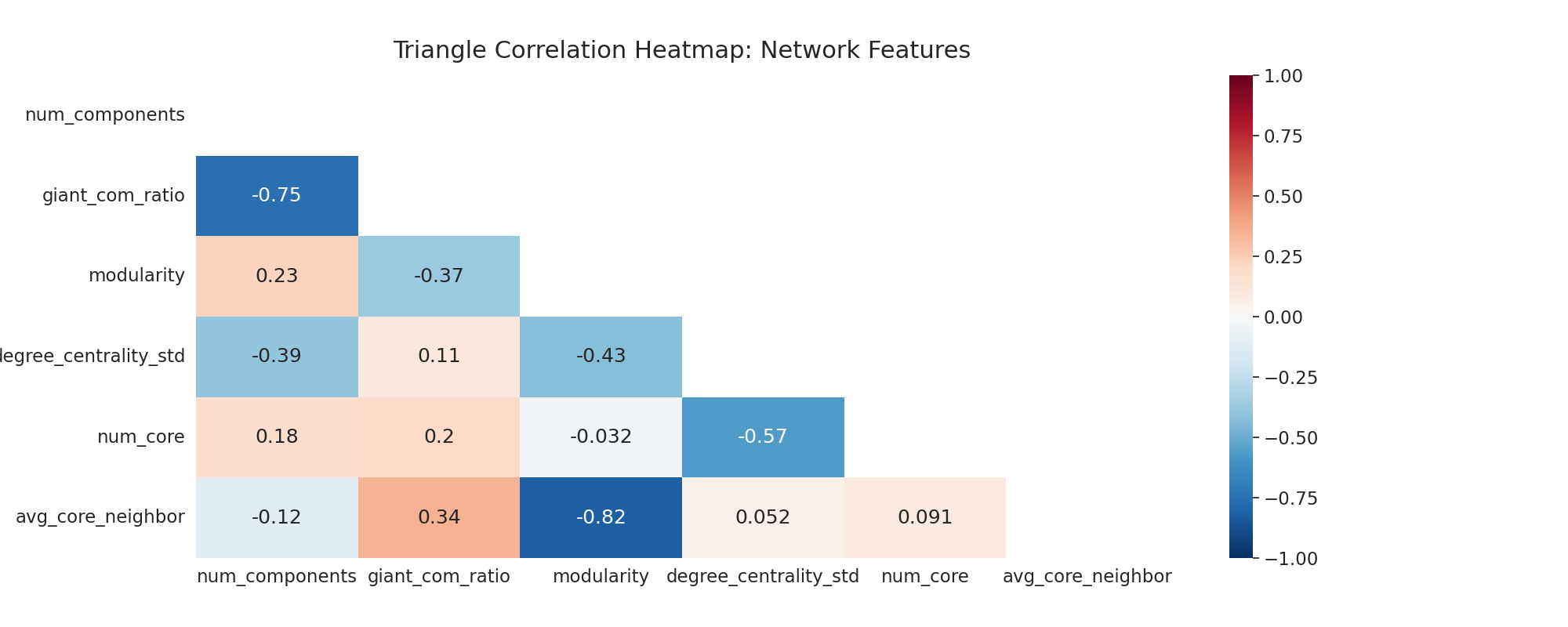} 
  \caption{Correlation heatmap of network features of the LUSD token.}
  \label{cor_LUSD}
\end{figure*}

\begin{figure*}[!htbp]
  \centering
  \includegraphics[width=\textwidth]{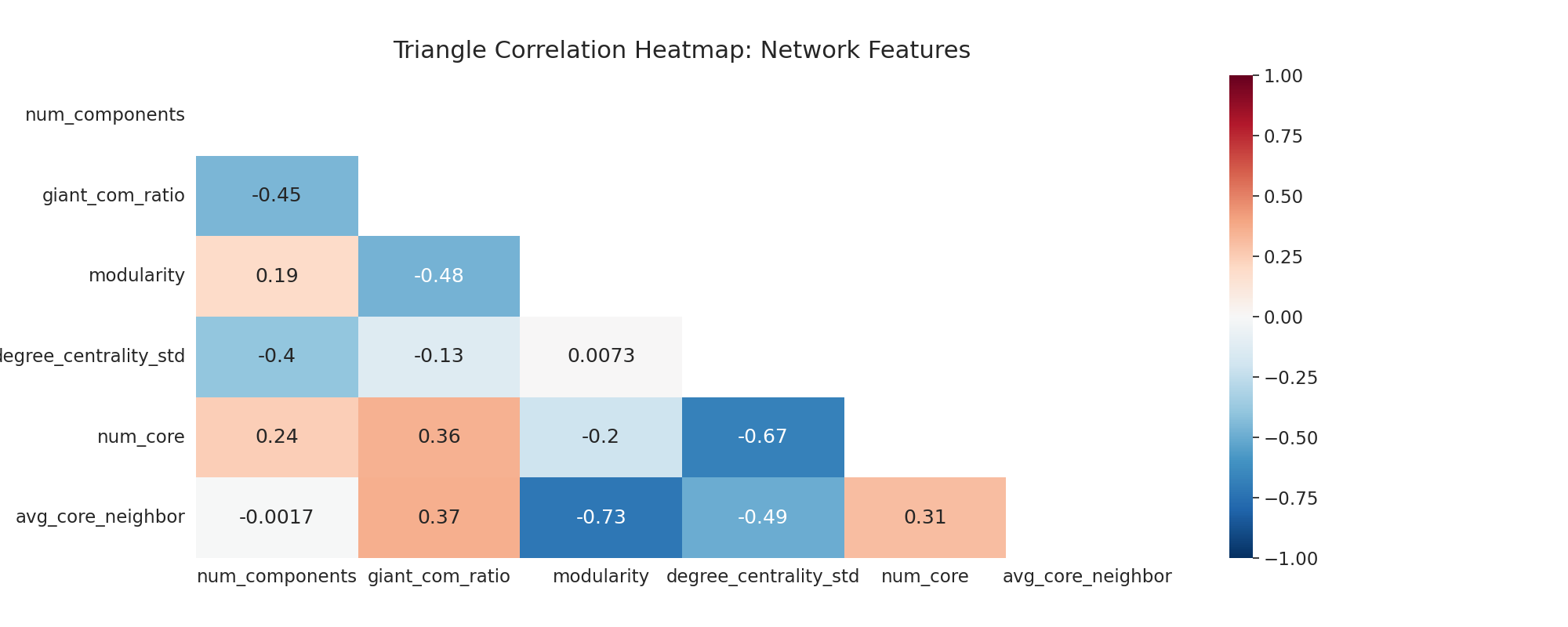} 
  \caption{Correlation heatmap of network features of the LQTY token.}
  \label{cor_LQTY}
\end{figure*}

\begin{figure*}[!htbp]
  \centering
  \includegraphics[width=\textwidth]{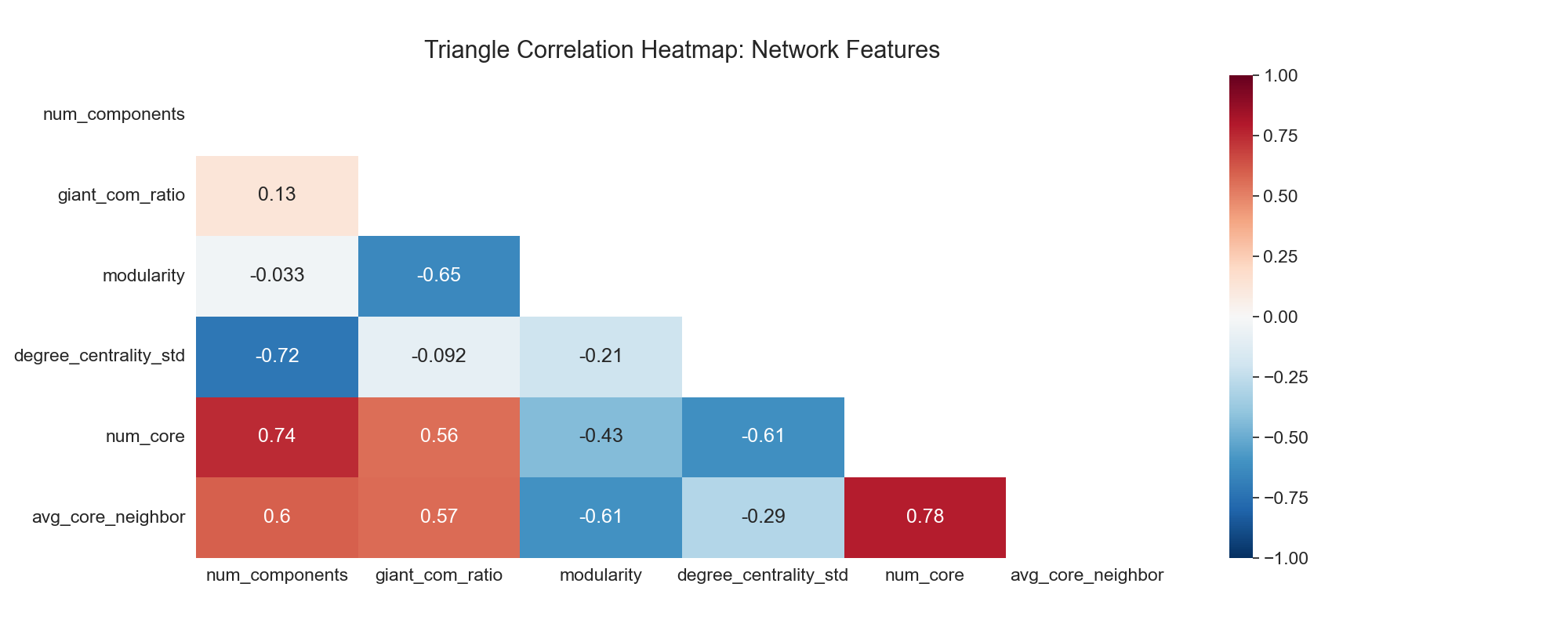} 
  \caption{Correlation heatmap of network features of the AAVE token.}
  \label{cor_AAVE}
\end{figure*}

\begin{figure*}[!htbp]
  \centering
  \includegraphics[width=\textwidth]{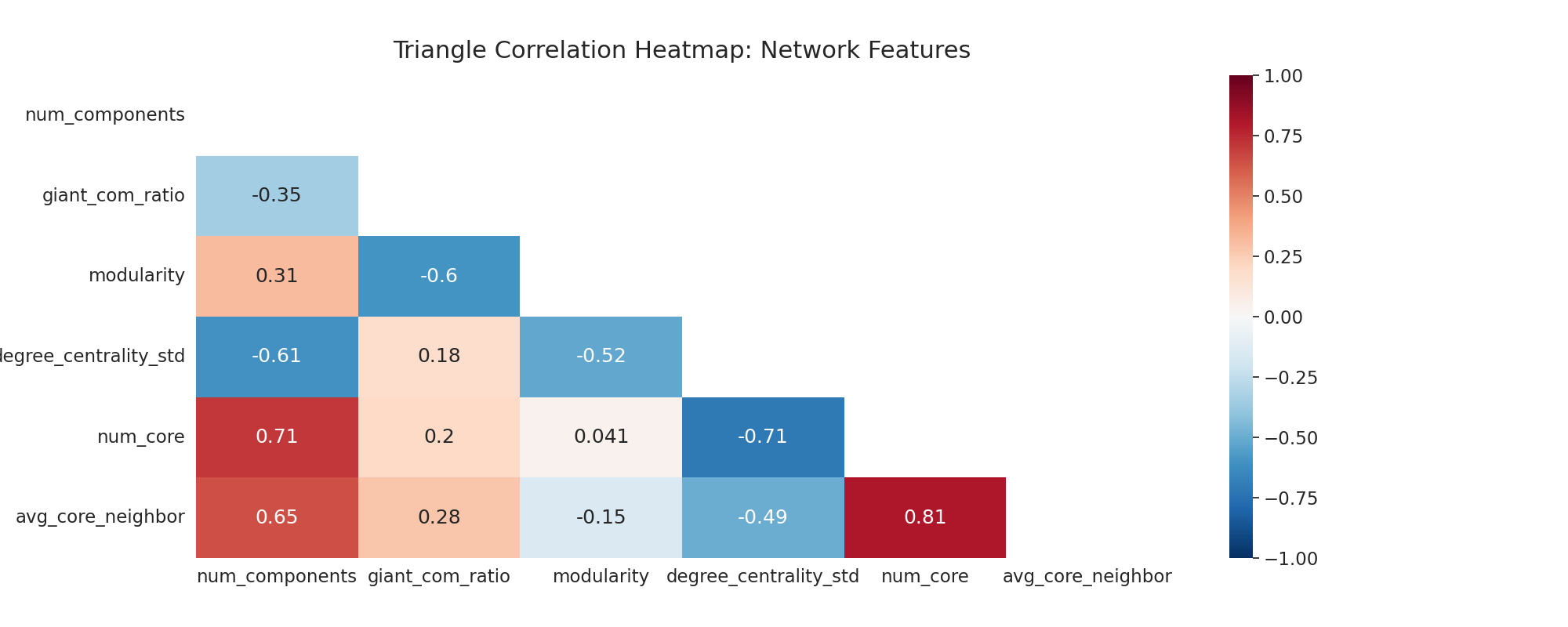} 
  \caption{Correlation heatmap of network features of the Dai token.}
  \label{cor_Dai}
\end{figure*}

\begin{figure*}[!htbp]
  \centering
  \includegraphics[width=\textwidth]{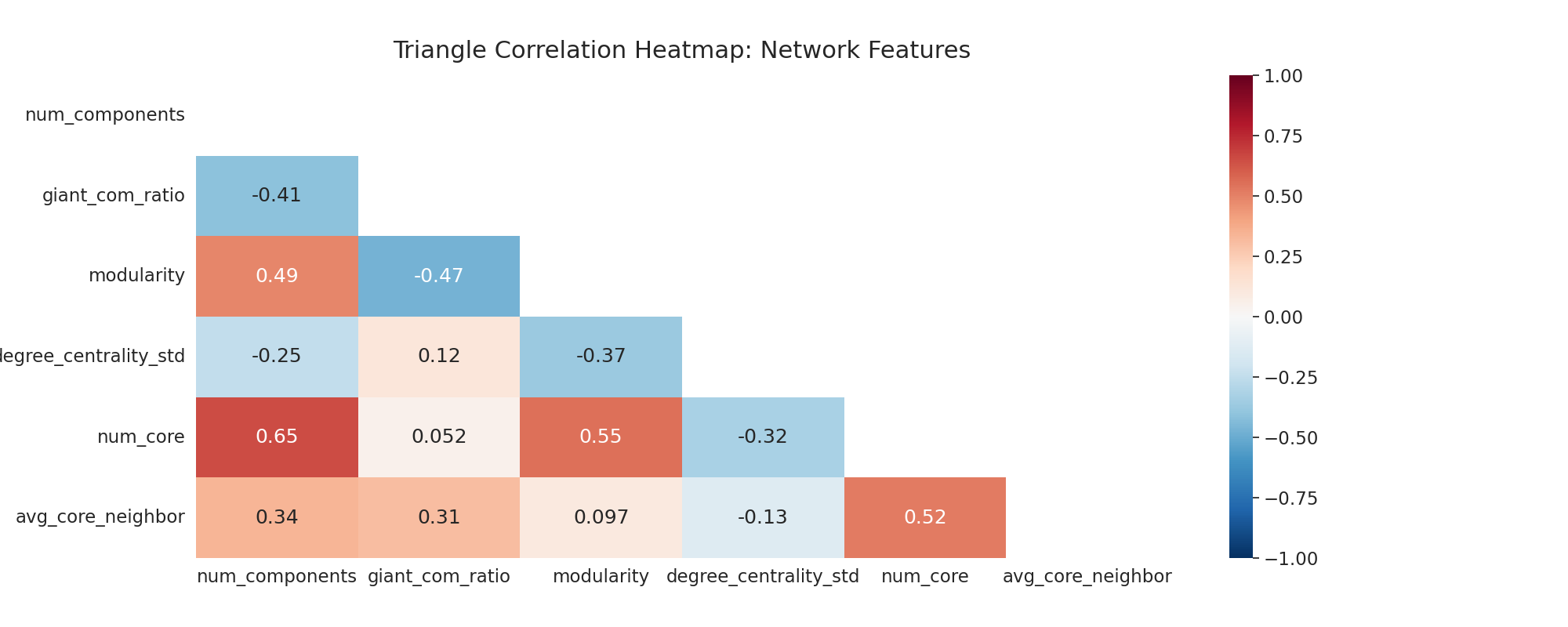} 
  \caption{Correlation heatmap of network features of the COMP token.}
  \label{cor_COMP}
\end{figure*}

\end{document}